%% file: main.tex
\documentclass[sigconf, screen, review]{acmart}

\usepackage{amsmath,amsfonts}
\usepackage{algorithmic}
\usepackage{graphicx}
\usepackage{textcomp}
\usepackage{xcolor}
\usepackage{multirow}
\usepackage{balance}

\usepackage{colortbl}

\newcommand{\ignore}[1]{}

\newcommand{\new}[1]{{\color{black}#1}}

\AtBeginDocument{%
  }

\setcopyright{none} 

\title{CODA: How to Mitigate ColumnDisturb\\ for (Almost) Free?}
\subtitle{\normalsize{Moinuddin Qureshi (Georgia Tech)}}

\input{abstract}

\setcopyright{none}
\renewcommand\footnotetextcopyrightpermission[1]{}

\begin{document}


\maketitle
\thispagestyle{plain}
\pagestyle{plain}

\input{intro}

\input{background}

\input{methodology}

\input{codae}

\input{codaf}

\input{codag}
\input{rega}

\input{related}

\input{conclusion}
\bibliographystyle{ACM-Reference-Format}
\bibliography{refs}
\end{document}

%% file: abstract.tex
\begin{abstract}

ColumnDisturb is a new data-disturbance error in which activations to an aggressor row cause bitflips in a victim row located hundreds of rows away (intra-subarray bitflips) and in victim rows in adjacent subarrays (inter-subarray bitflips).  Existing row-level Rowhammer defenses are unable to tolerate ColumnDisturb, as they perform mitigations in a small localized region around the aggressor row. Intra-subarray ColumnDisturb can be tolerated by solutions (such as SALT and REGA) that operate at subarray granularity.  However, to tolerate inter-subarray ColumnDisturb, such solutions must be extended with {\em ColumnDisturb Protection (CDP)}, which performs additional {\em Adjacent-Counter Increment (ACI)} for the neighboring subarrays.  The ACIs ensure that adjacent subarrays also undergo mitigation, even if they receive no demand activations.  Unfortunately, because ACIs occur at a 200\% rate relative to demand activations, they effectively increase the activations perceived by the bank to 3x, which causes significant slowdowns (17\% at a TRHD of 500) and refresh overheads. The goal of our paper is to tolerate ColumnDisturb while incurring negligible overheads. 

We propose {\em {CODA}}, a ColumnDisturb mitigation that significantly reduces the rate of ACI required to securely tolerate ColumnDisturb. We present three variants of CODA.  First, {\em CODA-E (Evade)}, which leverages the insight that ACI can be skipped if the neighboring subarray receives a demand activation, and reduces ACI by 2x. Second, {\em CODA-F (Fraction)}, which uses the timing duration of ColumnDisturb to do only a fractional increment for ACI, thereby reducing the rate of ACI by 2x-16x. Finally, {\em CODA-G (Gangskip)}, which is designed for solutions that operate at multi-subarray granularity and skips ACI for neighboring subarrays within the same gang, further reduces overall ACI by 2x-8x. Overall, CODA reduces ACI by 12x-1300x, thereby making it possible to tolerate ColumnDisturb while incurring zero performance and power overhead.

\end{abstract}


%% file: intro.tex
\ignore{
Flow
Rowhammer .. Tracking and mitigation (BR)
PRAC
Problem: BR-Attacks
SALT (Security)
Performance problem
SALT-C
Contributions

Takeaway

}




\begin{figure*}[!htb]
    \centering
\includegraphics[width=6.9in]{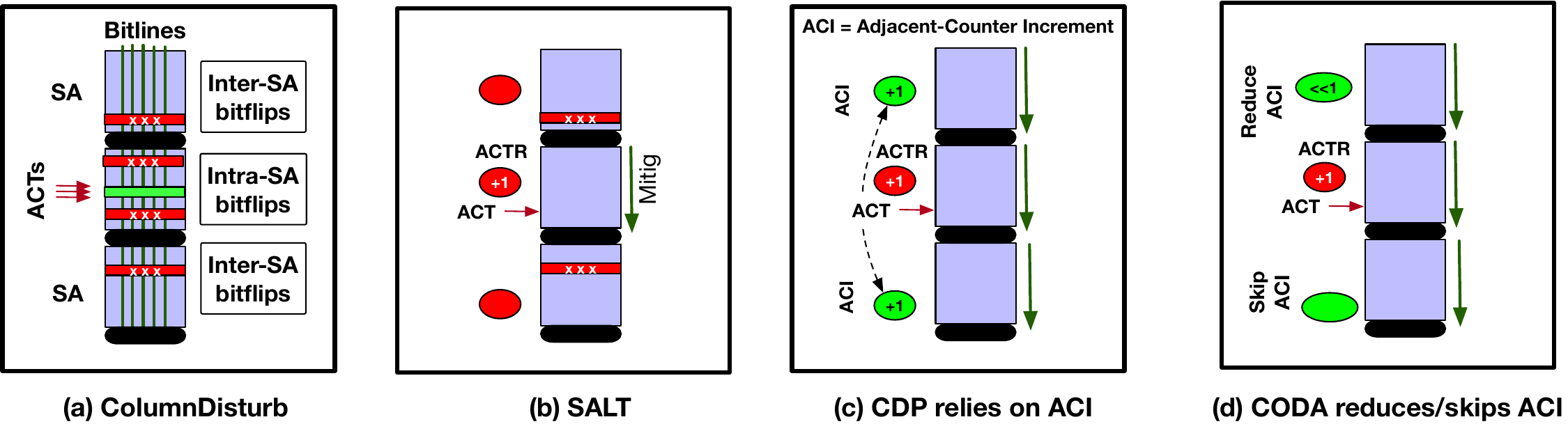}
    \caption{  (a) ColumnDisturb leverages shared bitlines to cause distant bitflips, both inter-subarray and intra-subarray. (b) Intra-subarray bitflips can be tolerated by SALT as it operates at subarray granularity. (c) SALT with {\em ColumnDisturb Protection (CDP)} can tolerate inter-subarray bitflips. On activation, CDP not only increments the ACTR of the demand subarray but also of the two adjacent subarrays. Such {\em Adjacent-Counter Increment (ACI)} increases the need for mitigation and incurs significant overheads (d). Our design, CODA, reduces ACI by 12x-1300x, thereby reducing the overheads of tolerating ColumnDisturb.    }
    \label{fig:intro}
\end{figure*}

\section{Introduction}

Data-Disturbance Errors (DDEs) violate the isolation property of memory systems by allowing access to one memory location to modify the data stored in another. DDEs are not only a reliability challenge but also a significant security threat, as an attacker can use DDEs to flip bits in critical data structures, such as the Page Tables, to escalate privileges.  As DRAMs scale to smaller sizes, we continue to encounter new forms of DDEs.  

The most well-known DDE vulnerability is Rowhammer~\cite{kim2014flipping}, in which frequent accesses to an aggressor row cause bitflips in nearby victim rows. The number of activations required to cause a bitflip is called the {\em Rowhammer Threshold (TRH)}. Typical mitigations for Rowhammer are based on tracking aggressor rows and refreshing a small number of victim rows on either side of the aggressor row.  RowPress~\cite{rowpress} is another recent DDE, in which the aggressor row is kept open for a long time, thereby causing charge leakage in proportion to the row-open time.  As RowPress reduces the number of activations required to induce a bitflip in the nearby victim rows, it reduces the effective TRH.  Fortunately, RowPress can be easily tolerated with existing hardware-based Rowhammer solutions by converting the row open-time into equivalent activations~\cite{saxena2024impress}.

A recent paper introduced {\em ColumnDisturb}~\cite{columndisturb}, a new class of disturbance errors, whereby activations in aggressor row can cause bitflips in distant victim rows, which are located hundreds of rows away from the aggressor row within the same subarray (we call this {\em intra-subarray ColumnDisturb}) or in victim-rows located in adjacent subarrays (we call this {\em inter-subarray ColumnDisturb}). ColumnDisturb exploits the fact that in open-bitline architectures, adjacent subarrays share a sense amplifier and that bitlines run across the adjacent subarrays, as shown in Figure~\ref{fig:intro}(a).  This sharing of bitlines breaks spatial isolation between subarrays, allowing an activation in one subarray to affect the data stored in another. The pattern of ColumnDisturb is similar to Rowhammer and RowPress, just that the time duration for ColumnDisturb is in the range of several milliseconds.  Existing Rowhammer solutions that perform row-granularity tracking and mitigations cannot tolerate ColumnDisturb as they refresh only a small number of rows (typically 1-2) on either side of the aggressor row.  

ColumnDisturb can be mitigated by Rowhammer mitigations that operate at a subarray granularity, such as SALT~\cite{salt}. As shown in Figure~\ref{fig:intro}(b),  SALT provisions each subarray with an {\em Activation Counter (ACTR)}, which is incremented on each activation to the subarray. When the activation counter reaches a specific value, SALT performs a sequential refresh of several rows within the subarray. The mitigation rate of SALT is designed to ensure that SALT refreshes all rows within the subarray before the subarray receives TRH activations. As SALT provides Blast-Radius freedom (the victim row can be anywhere within the same subarray as the aggressor), SALT implicitly handles intra-subarray ColumnDisturb. However, SALT cannot handle inter-subarray ColumnDisturb. 

To tolerate inter-subarray ColumnDisturb, SALT can be provisioned with {\em ColumnDisturb Protection (CDP)}~\cite{salt}, whereby each demand activation to a subarray also performs additional {\em Adjacent-Counter Increments (ACI)} for the two adjacent subarrays, as shown in Figure~\ref{fig:intro}(c). ACIs enable adjacent subarrays to perform mitigation even if they receive no demand activations, thereby providing protection against inter-subarray ColumnDisturb. As CDP issues two ACIs per demand activation, it significantly increases the mitigation requirements of SALT (equivalent to a bank incurring 3x activations), resulting in large performance and power overheads (e.g., CDP increases SALT's slowdown from 0.3\% to 17\% at TRHD of 500). The goal of our paper is to mitigate ColumnDisturb while incurring negligible performance and power overheads.

The key insight of our paper is to reduce the overhead of ColumnDisturb protection by reducing the number of ACIs issued per demand activations.  We propose {\em CODA ({\underline{Co}}lumn{\underline{D}}isturb Mitigation with Reduced {\underline{A}}CIs)}, as shown in Figure~\ref{fig:intro}(d). We present three variants of CODA, each targeting a different inefficiency in ACI. 

Our first design, {\em CODA-E (Evade)}, avoids the ACI  to the neighboring subarray if the neighboring subarray receives a demand activation. The key observation of CODA-E is that if a neighbor receives sufficient demand activations, it will undergo mitigations anyway due to the demand activity and does not require ACI for mitigations.  We show, both theoretically and experimentally, that CODA-E can halve the rate of ACIs, reducing it from 200\% to 100\%. 

Our second design, {\em CODA-F (Fraction)}, exploits the fact that ColumnDisturb is a much slower attack compared to Rowhammer and lasts for several milliseconds. For example, the ColumnDisturb paper targets solutions for a ColumnDisturb attack duration of 8 ms.  As Rowhammer solutions are designed to tolerate much faster attacks, we should lower the rate of ACI so that the neighboring subarray can fully refresh its rows within our target ColumnDisturb duration.  For example, if we assume that a row-open time of 500 ns is equivalent to 1 activation, then the attacker must inflict 16K equivalent activations within 8 ms to cause ColumnDisturb.  If the Rowhammer solution is designed to refresh all rows in the subarray within 1K activations, then it would be sufficient to issue one ACI every 32 demand activations, rather than 1 per demand activation. CODA-F reduces the ACI rates by 16x-2x for TRH of 500 to 4K, respectively. Thus, CODA-F is especially attractive for lower TRH.  CODA-E and CODA-F are synergistic and can be combined to substantially reduce the rate of ACI from 200\% for CDP to 1.4\% (TRH of 500) or 17\% (TRH of 4K).

Our third design, {\em CODA-G (Gangskip)}, is designed for solutions that operate at multi-subarray granularity. For example, {\em Ganged-SALT}~\cite{salt} is a variant of SALT that operates at the granularity of 2 (at TRH of 1K) to 8 (at TRH of 4K) subarrays. As the gang shares the same activation counter and metadata, Ganged-SALT reduces the SRAM storage overhead by 2x-8x. The key observation in CODA-G is to skip the ACI for the adjacent subarray when that subarray is within the same gang, since gang-based mitigation will also refresh all rows in the adjacent subarray.  CODA-G reduces the rate of ACI by 2x (TRH of 1K) to 8x (TRH of 4K) and becomes even more attractive at higher TRH.  CODA-G can be combined with CODA-E and CODA-F to reduce the rate of ACI from 200\% (for CDP) to 0.23\% (at TRH of 1K) or 0.16\% (at TRH of 4K).  

As CODA virtually eliminates ACIs, CODA makes it practical to tolerate ColumnDisturb with the same solution as Rowhammer, while incurring negligible performance and power overheads (0.01\%).  The storage overhead of CODA is small (4 bits per subarray).


The principles of CODA are not specific to SALT and are applicable to any Rowhammer mitigation that operates at a subarray granularity.  For example, we also analyze ColumnDisturb mitigation with REGA~\cite{REGA_SP23}. REGA tolerates Rowhammer by provisioning the subarray with an extra row buffer and generating a refresh of one or more rows within a subarray on each demand activation. REGA can protect intra-subarray ColumnDisturb; however, to protect inter-subarray ColumnDisturb, it will also need ACI for neighboring subarrays.  A design based on CDP (REGA-CDP) will issue two ACI for each demand activation.  REGA-CDP incurs 3x the refresh-power overheads as REGA for mitigation. When REGA is implemented with CODA (REGA-CODA), the ACIs are reduced by 12x-140x, thereby virtually eliminating the increase in refresh-power overhead required to tolerate ColumnDisturb. 




\vspace{0.05 in}

\noindent{\bf Contributions:} Our paper makes the following contributions: 

\begin{enumerate}
    \item We show that the main overhead of tolerating ColumnDisturb is due to {\em Adjacent-Counter Increment} (ACI), where each activation increments the counter for three subarrays. 
    
    \item We propose {\em CODA}, which significantly reduces the rate of ACI by leveraging demand-activations to skip ACI (CODA-E) and performing fractional increments for ACI (CODA-F). 

    \item We propose {\em CODA-G}, which is tailored to mitigations that operate at multi-subarray granularity and skips ACI when the adjacent subarray is within the same gang.  
\end{enumerate}

As CODA reduces the rate of ACI by 12x-1300x, thereby making it practical to tolerate ColumnDisturb at nearly zero overhead.

%% file: background.tex
\ignore{
Threat Model
DRAM, contains subarray 
(timings) and refresh

Rowhammer: Space and Time

Mitigation and Blast Radius Example (picture from prior papers)
Goal
}


\section{Background and Motivation}

\subsection{Threat Model}
Our threat model assumes that an attacker can issue memory requests for arbitrary addresses. The attacker knows the defense algorithm and any relevant parameters. We declare an attack successful when a row incurs sufficient charge loss to trigger a bit flip. We assume that the attacker uses the most effective pattern \new{(including attacks that spread activations across different subarrays)} based on the defenses employed.  We assume that the inter-subarray bitflips from ColumnDisturb are restricted to only the adjacent subarrays. We further assume that any row within a subarray can be used to induce ColumnDisturb in adjacent subarrays, allowing the attacker to spread activations across many rows within a subarray. 

\subsection{DRAM Architecture and Parameters.}

\begin{figure}[!htb]
    \centering
\includegraphics[width=2in]{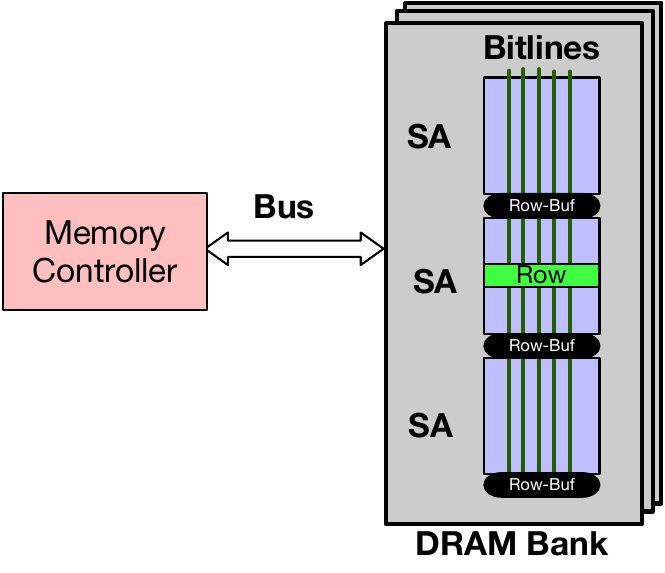 }
    \caption{DRAM Architecture. A bank consists of subarrays. Open Bitline Architectures shares row-buffer between adjacent subarrays, and bitlines run across multiple subarrays.}
    \label{fig:dram}
\end{figure}

DRAM chips are organized into arrays comprising {\em rows} and {\em columns}.  To access the data in DRAM, the row is accessed using the {\em word-line}, and the charge on the DRAM cells is sensed using a sense amplifier, and the data is stored in a {\em Row-Buffer (RB)}. The DRAM chip is divided into a number of banks (32 for DDR5), with only one row-buffer per bank architecturally visible to the Memory Controller. However, internally, each bank consists of smaller independent units called the {\em Subarray}~\cite{SALP,REGA_SP23,columndisturb,salt,smd,mirza,AutoRFM}, each equipped with a row-buffer and sensing circuit.  We assume an open bitline architecture where adjacent subarrays share a row-buffer, and the bitlines are spread over multiple subarrays, as shown in Figure~\ref{fig:dram}.  Each subarray typically contains 512 rows.  In our system, each DRAM bank contains 128K rows, so each bank has 256 subarrays. 


To access data from DRAM, the memory controller must first issue an activation (ACT) to open the row. To access data from another conflicting row, the opened row must be precharged (PRE) first. To ensure data retention, all data in the DRAM must be refreshed within the retention period (tREFW of 32ms). To reduce the latency of refresh, memory is divided into 8192 groups, and a REF command, issued every tREFI (3.9 microseconds), refreshes one group. As our bank contains 128K rows, each REF must refresh 16 rows in the bank. We assume that consecutive rows in a bank are spread to consecutive subarrays, so that accesses to consecutive rows are not focused within the same subarray~\cite{SALP,mirza}. 


\ignore{
\begin{table}[!htb]
  \centering
  \vspace{-0.1in}
  \caption{\new{DRAM Timings (DDR5-6000AN~\cite{JEDEC-PRAC}).}}
  \begin{footnotesize}
  \label{table:Params}
  \begin{tabular}{lcc}
    \hline
    \textbf{Parameter} & \textbf{Description} & \textbf{Baseline} \\ \hline \hline

    tRCD     & Time for performing ACT & 14 ns \\ 
    tRP     & Time to precharge an open row & 14 ns  \\ 
    tRAS     & Min. time a row must be kept open & 32 ns \\ 
     tRC     & Time between two ACTs to bank & 46 ns  \\ \hline 
        
    tREFW     & Refresh Period & 32 ms \\ 
    tREFI     & Time between two REF Commands & 3900 ns  \\ 
    tRFC      & Execution Time for REF Command & 410 ns  \\ 

\hline 
  \end{tabular}
  \end{footnotesize}
\end{table}


}

\subsection{Problem of Data-Disturbance Errors}

As DRAM cells scale down, the inter-cell distance decreases, and activity in one cell starts to affect the data stored in another cell. This is called a {\em Data-Disturbance Error (DDE).} The most well-known DDE is {\em Rowhammer}~\cite{kim2014flipping}, which was discovered more than a decade ago.  Rowhammer occurs when an aggressor row is repeatedly activated, causing a small amount of charge loss in the cells of the neighboring row. When this charge loss exceeds a given threshold, it induces bit-flips in neighboring rows.  The minimum number of activations to an aggressor row to cause a bit-flip in a victim row is called the {\em Rowhammer Threshold (TRH)}. TRH is reported either for a single-sided pattern {\em (TRHS)} or a double-sided pattern {\em (TRHD)}. TRH has dropped from 139K (TRHS) in 2014~\cite{kim2014flipping} to 4.8K (TRHD) in 2020~\cite{kim2020revisitingRH}. Hardware solutions for mitigating Rowhammer typically rely on a tracking mechanism to identify the aggressor rows, and then perform mitigation by refreshing a small number of victim rows on either side of the aggressor row (as specified by the {\em{Blast-Radius}}). 

As DRAM scales down, we encounter new DDE modalities. For example, about three years ago, {\em RowPress}~\cite{rowpress} was released as a new form of DDE. With RowPress, the aggressor row is kept open for a long time.  The open row causes a small amount of leakage on the bitlines, and the longer the row remains open, the greater the leakage.  For example, keeping the row open for 500 ns results in twice as much leakage as activating and immediately precharging the row. As RowPress causes greater leakage per activation than Rowhammer, it reduces the effective TRH required to induce a bitflip in neighboring rows. RowPress can be handled by existing Rowhammer mitigation by simply converting the row open time into equivalent activations~\cite{saxena2024impress}, so a row that is open for a long time is treated as having induced more activations.  This ability to solve new DDEs using existing solutions is useful, as it avoids the complexity of maintaining separate solutions for different DDEs.

\subsection{ColumnDisturb: New Modality of DDE}

Both Rowhammer and RowPress affect victim rows that are within a close spatial proximity of the aggressor row.  A few months ago, a new DDE vulnerability, ColumnDisturb~\cite{columndisturb}, was revealed that can cause bitflips in victim rows hundreds of rows away from the aggressor row.  The key property used in ColumnDisturb is the shared bitlines that span many rows and subarrays, which cause activity in one row to affect leakage in distant rows, even when the victim rows are in another subarray.  The most effective pattern for ColumnDisturb is similar to RowPress (keeping a row open for a long time and repeating it), with the caveat that ColumnDisturb is much slower and must be repeated for several milliseconds~\cite{columndisturb}.  

We classify the bitflips caused by ColumnDisturb into two categories: First, {\em Intra-Subarray ColumnDisturb}, where the distant victim row remains within the same subarray as the aggressor row.  Second, {\em Inter-Subarray ColumnDisturb}, where the victim row is in the subarray adjacent to the subarray of the aggressor row (experiments show that ColumnDisturb affects only the adjacent subarray).  ColumnDisturb can be mitigated by reducing the refresh rate; however, this incurs significant performance and power overheads. For example, to tolerate a ColumnDisturb of 8ms, we would need to increase the refresh rate by a factor of 4 (from 32ms to 8ms), which would cause a 22\% slowdown and quadruple the refresh power. 

Existing row-granularity Rowhammer solutions cannot tolerate ColumnDisturb, as they refresh only a few nearby victim rows.  To tolerate ColumnDisturb in a practical manner, we focus on solutions, such as SALT~\cite{salt} and REGA~\cite{REGA_SP23}, that operate at subarray granularity. Without loss of generality, we focus on ColumnDisturb mitigation using SALT (we discuss REGA in detail in Section~\ref{sec:rega}).

\subsection{SALT}

\begin{figure}[!htb]
    \centering
\includegraphics[width=2.5in]{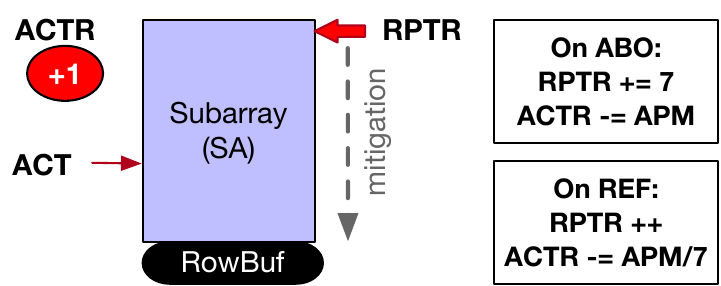 }
    \caption{Overview of SALT with Refresh Coordination. SALT provisions ACTR and RPTR with each subarray (or gang).}
   \vspace{-0.1 in}
    \label{fig:salt}
\end{figure}

SALT~\cite{salt} is a recent Rowhammer mitigation that uses subarray-level tracking and mitigation to handle bitflips beyond the Blast-Radius. Figure~\ref{fig:salt} shows an overview of SALT. SALT equips each subarray with an {\em Activation Counter (ACTR)} and a {\em Refresh Pointer (RPTR)}.  On each activation, the ACTR of the accessed subarray is incremented.  When ACTR reaches a specified value, SALT initiates an {\em Alert-Back-Off (ABO)}~\cite{JEDEC-PRAC} signal to obtain the time for mitigation.  The time of ABO is sufficient to refresh 7 rows, so 7 rows starting with RPTR are refreshed, and the RPTR is incremented by 7.  The ACTR is decremented by {\em APM (Activations Per Mitigation)}.

The APM value dictates the threshold tolerated by SALT. For example, with an APM of 13, SALT can tolerate a TRHD of 500 (1K activations across all rows in the subarray).  This would mean issuing an ABO every 13 activations, which incurs significant performance overhead (e.g., 17\% at a TRHD of 500). SALT reduces ABO overheads by leveraging the time-based refresh (REF) to avoid activity-based refreshes.  So, if the activity can be handled by REF, then ABO is avoided.  This feature, called {\em Refresh Coordination}, is vital for keeping SALT's performance overhead low.  For example, Refresh Coordination reduces the slowdown of SALT from 17\% to 0.3\% at a TRHD of 500, and to 0\% at a TRHD of 1K and beyond. {\bf We assume SALT is always implemented with Refresh Coordination}.

As the slowdown of SALT is 0\% for TRHD $\ge$ 1K, we can reduce the storage overhead of SALT by operating at a multi-subarray granularity.  Such a {\em Ganged-SALT}~\cite{salt} design forms a gang of 2/4/8 subarrays for TRHD of 1K/2K/4K, respectively.  As Ganged-SALT uses one ACTR and one RPTR per gang, it reduces storage overhead by 2x-8x.  For example, Ganged-SALT can tolerate TRHD of 4K with only 72 bytes of SRAM per bank (which is lower than some TRR implementations already deployed in DDR4~\cite{ProTRR,jattke2021blacksmith}).

\subsection{ColumnDisturb Protection with SALT}

As SALT operates at a subarray granularity, it implicitly tolerates intra-subarray ColumnDisturb, as SALT refreshes all the rows within the subarray before 2*TRHD activations are inflicted on the subarray.  However, SALT cannot tolerate inter-subarray ColumnDisturb. For example, if the attacker focuses activations on a single subarray, ColumnDisturb can still cause failures in neighboring subarrays because adjacent subarrays do not get mitigated by SALT.


To mitigate inter-subarray ColumnDisturb, SALT can be equipped with a specific extension called {\em ColumnDisturb Protection (CDP)}~\cite{salt}.  With CDP, each demand activation not only increments the ACTR of the given subarray but also performs {\em Adjacent-Counter Increment (ACI)} for the two adjoining subarrays.  The role of ACI is simply to increment the counter of the neighboring subarray without doing any activation.  Thus, CDP increments 3 activation counters per demand activation. This ensures that even if activations are concentrated on a single subarray (e.g., SA-2), the neighboring subarrays (e.g., SA-1 and SA-3) are forced to undergo mitigation, and hence they are able to tolerate inter-subarray ColumnDisturb.

\begin{figure}[!htb]
    \centering
\includegraphics[width=1.5in]{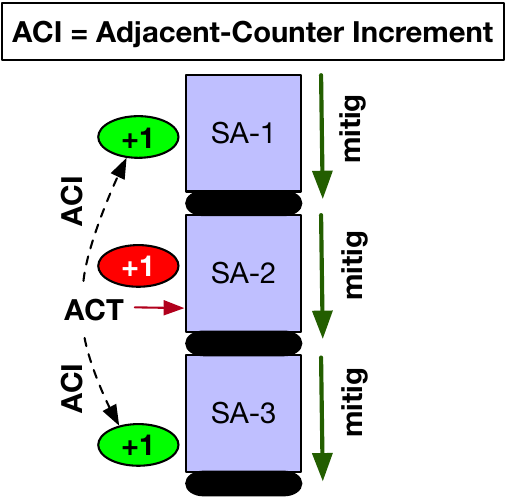 }
    \caption{Overview of ColumnDisturb Protection (CDP).  On activation, CDP issues ACI to neighboring subarrays. }
   \vspace{-0.1 in}
    \label{fig:cdp}
\end{figure}

\subsection{Overheads of CDP}

As CDP causes two ACIs per activation, it requires 3x the mitigation resources as SALT.  This causes significant overhead for SALT and Ganged-SALT.  Table~\ref{tab:cdpperf} shows the average slowdown of SALT and Ganged-SALT, both with and without CDP, as TRHD is varied from 500 to 4K. Note that SALT handles 2*TRHD activations at {\em arbitrary} locations within the subarray (e.g. focused on a single row or spread across multiple rows). CDP incurs significant overhead, increasing the slowdown of SALT/Ganged-SALT from 0.3\% to 17\%.   We want to tolerate ColumnDisturb while avoiding these overheads.


\begin{table}[htb]
\centering
\begin{small}
\caption{Avg. Slowdown of CDP for SALT and Ganged-SALT}
\vspace{-0.05 in}
\label{tab:cdpperf}
\begin{tabular}{ccccc}
\toprule
\textbf{TRHD} & \textbf{500} & \textbf{1K} & \textbf{2K} & \textbf{4K} \\ \hline
\midrule
SALT                & 0.3\% & 0\% & 0\% & 0\% \\
SALT (CDP)          & {\bf 17.8\%} &  {\bf1.9\%} & 0\% & 0\% \\ \hline 
Ganged-SALT         & N/A &  0.3\% & 0.3\% & 0.3\% \\
Ganged-SALT (CDP)   & N/A & {\bf17.7\%} & {\bf17.3\%} & {\bf 16.8\%} \\
\bottomrule
\end{tabular}
\vspace{-0.05 in}

\end{small}
\end{table}

\subsection{Goal of Our Paper}

The goal of our paper is to tolerate ColumnDisturb while incurring negligible overheads.  Our key observation is that the primary source of overhead for ColumnDisturb mitigation is the ACI issued at every activation (occurring at a rate of two per activation, i.e., 200\%).  If we reduce the rate of ACI, we can enable low-overhead ColumnDisturb mitigations.  To that end, we propose {\em CODA, a ColumnDisturb Mitigation with Reduced-ACI}. We first present our experimental methodology before presenting our solution.

%% file: methodology.tex
\clearpage

\section{Experimental Methodology}

To ensure consistency with prior work, we use the publicly available artifact~\cite{salt} from the SALT paper for our evaluations.  The artifact uses Memsim~\cite{qureshi2024mint,qureshi2024moat,salt}, a cycle-level multi-core simulator with a detailed memory model. Table~\ref{table:system_config} shows our configuration. We use the updated DDR5 timing specifications. We used a minimalist open-page mapping~\cite{kaseridis2011minimalist}.  We use an {\em adaptive} paging policy that closes the page if there are no pending requests to the opened row. Our adaptive policy outperforms both open-page and closed-page.  


For SALT, we use APM of 13/26/53/106 (and ATH of 2x of APM) for TRHD of 500/1K/2K/4K, respectively.  We assume SALT always uses Refresh-Coordination.

\begin {table}[htb]
\begin{footnotesize}
\begin{center} 
\vspace{-0.1in}
\caption{Baseline System Configuration}
\vspace{-0.1in}
\begin{tabular}{|c|c|}
\hline
  Out-of-Order Cores           & 8 core, 4GHz, 4-wide, 256 entry ROB   \\
  Last Level Cache (Shared)    & 8MB, 16-Way, 64B lines \\ \hline
  Memory specs                 & 32 GB, DDR5  \\
  t${ALERT}$                   & 180ns (normal) + 350ns (RFM) = 530ns\\
  Banks x Sub-channel x Rank   & 32$\times$2$\times$1 \\
  Rows                         & 64K rows per bank, 8KB rows\\ \hline
  Mapping and Closure Policy   & Minimalist Mapping, \new{Adaptive Page Closure} \\ \hline
\end{tabular}
\vspace{-0.1in}
\label{table:system_config}
\end{center}
\end{footnotesize}
\end{table}

We use 13 benchmarks from SPEC-2017 with at least one L3-Miss per 1K instructions (L3-MPKI), six from GAP~\cite{GAP}, four from STREAM~\cite{mccalpin:memory}, and two data-analytics benchmarks (KMeans~\cite{lloyd:least} and MassTree~\cite{mao:cache}).  We run the workloads in 8-core rate-mode, until each core completes 1 billion instructions (representative slice). We measure performance using weighted speedup. Table~\ref{table:workloads} shows workload characteristics, including L3-MPKI and ACT-per-tREFI (per bank).  Our workloads are memory-intensive. 


\begin{table}[ht]
\centering
\begin{small}
\vspace{-0.05 in}
\caption{Workload Characteristics}
\vspace{-0.1 in}
\label{table:workloads}
\begin{tabular}{|c|l||c|c|}
\hline 
\textbf{Suite} & \textbf{Benchmark} & \textbf{L3-MPKI} & \textbf{ACT-per-tREFI} \\
 & &  & \textbf{ (per Bank) } \\
\hline \hline
\multirow{13}{*}{SPEC2K17} 
  & bwaves      & 42.7 & 17.7 \\
  & fotonik3d   & 28.3 & 26.0 \\
  & lbm         & 26.7 & 27.2 \\
  & parest      & 23.2 & 15.6 \\
  & mcf         & 23.1 & 18.1 \\
  & roms        & 11.6 & 16.6 \\
  & omnetpp     &  9.3 & 23.4 \\
  & xz          &  5.1 & 23.6 \\
  & cam4        &  5.0 & 13.9 \\
  & cactuBSSN   &  3.5 & 16.3 \\
  & wrf         &  1.2 &  4.6 \\
  & xalancbmk   &  1.1 &  4.7 \\
  & blender     &  1.0 &  5.2 \\
\hline
\multirow{6}{*}{GAP}
  & ConnComp    & 86.2 & 27.6 \\
  & PageRank    & 46.4 & 21.5 \\
  & TriCount    & 52.2 & 11.2 \\
  & BFS         & 37.8 & 19.0 \\
  & BC          & 20.7 & 14.2 \\
  & SSSPPath    & 10.3 & 12.5 \\
\hline
\multirow{4}{*}{STREAM}
  & add         & 15.6 & 14.2 \\
  & triad       & 13.4 & 14.0 \\
  & copy        & 12.5 & 13.0 \\
  & scale       & 10.4 & 12.7 \\
\hline
\multirow{2}{*}{ANALYTICS}
  & kmeans      &  8.9 & 18.6 \\
  & masstree    &  4.8 & 15.8 \\
\hline \hline 
Average & & 21.6 & 16.7 \\ \hline
\end{tabular}
\vspace{-0.1 in}
\end{small}
\end{table}

%% file: codae.tex
\section{CODA-E: Evade ACI on Demand Activation}

The main source of overhead in ColumnDisturb mitigation is the {\em Adjacent-Counter Increment (ACI)} for neighboring subarrays. Thus, each demand activation increments three counters (one for demand and two ACIs for neighbors).  We propose {\em CODA} to reduce the overhead of mitigating ColumnDisturb by either skipping or reducing the ACI to a much lower rate.  We propose three variants of CODA: {\em CODA-E (Evade)}, {\em CODA-F (Fraction)}, and {\em CODA-G (Gangskip)}.  Each variant targets a different inefficiency in ACI.  In this section, we focus on CODA-E.  

\subsection{Overview and Design of CODA-E}

CODA-E is based on the observation that the ACI-based increment of neighboring counters is needed only in cases where the underlying rate of mitigation of the neighboring subarrays would not be enough to refresh all the rows in those subarrays. This can occur in pathological cases, where the attacker focuses all activations on a single subarray, so that neighboring subarrays receive no activations (and thus cannot undergo any mitigation). However, for benign workloads, activations are spread across subarrays, so neighboring subarrays also receive demand activations and undergo mitigations.  This rate may be sufficient to refresh the neighboring subarray without requiring an ACI-based increment. The key insight for CODA-E is to skip the ACI-based increment if the subarray receives a demand increment.  

\begin{figure}[!htb]
    \centering
\includegraphics[width=1.6 in]{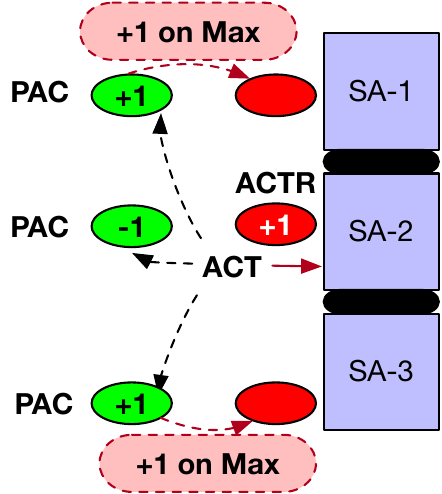 }
    \caption{Overview of CODA-E.  CODA-E adds a Pending-Activation Counter (PAC) to allow overriding of the ACI-based increment by a later demand ACT for that subarray. }
   \vspace{-0.1 in}
    \label{fig:codae}
\end{figure}

To implement CODA-E, we need to delay the ACI-based increment so that the subarray has time to override it with a demand activation.  Figure~\ref{fig:codae} shows an overview of our CODA-E design.

\begin{figure*}[!htb]
    \centering
\includegraphics[width=7.2in]{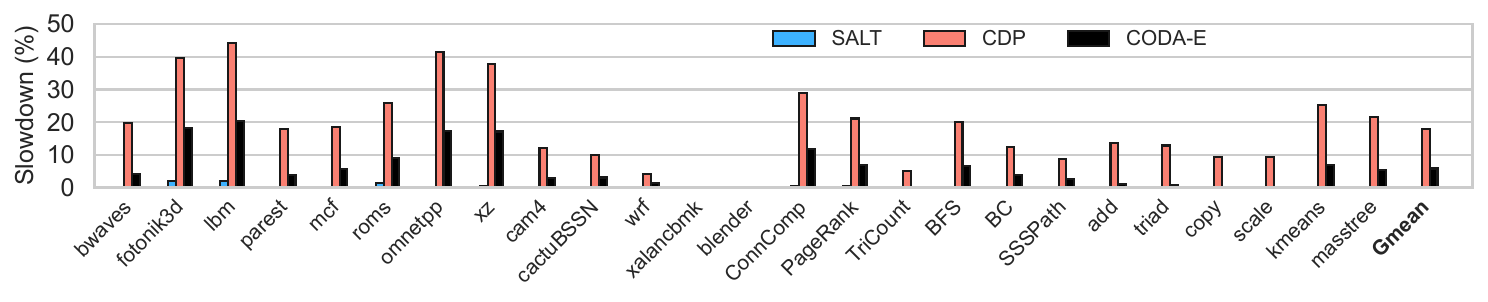}

\vspace{-0.15 in}
    \caption{Slowdown of SALT, CDP, and CODA-E at TRHD=500 compared to unprotected baseline. The average (Geometric mean) slowdown of SALT (without any ColumnDisturb mitigation) is 0.3\%, SALT with CDP is 17.8\%, and SALT with CODA-E is 5.9\%.}
\vspace{-0.1 in}
    \label{fig:codaeperf}
\end{figure*}

To allow flexibility for ACI-based increment to be later overridden by a demand ACT, CODA-E provisions each subarray with a separate auxiliary counter, called {\em PAC (Pending-Activation Counter)}.  On a demand activation to a subarray, the ACTR associated with the subarray is incremented. Instead of incrementing the ACTR of the two neighboring subarrays, CODA-E increments its PAC. Further, as the subarray receiving the demand ACT already has activity that could trigger mitigation, we reduce the PAC associated with the demand subarray (if it is non-zero).  Thus, in steady state, each demand ACT increments two PACs and decrements one PAC, so the total amount of PAC increments per ACT is limited to one. 

If an increment to PAC would bring it to its maximum value (determined by the number of bits allocated to the PAC counter), we bypass PAC and increment ACTR directly. Having more bits for PAC increases the time between PAC increments and the ability of a later demand ACT to reduce PAC for that subarray. However, increasing the number of bits also increases the storage cost and the number of pending increments, which can affect the tolerated thresholds. Unless specified otherwise, we use a 4-bit PAC.

\subsection{Impact on Rate of ACI}

We note that the theoretical lower bound for CODA-E for ACI is 100\%, which is much lower than the 200\% for CDP.  This is because, on each demand ACT, we cause two PAC increments and reduce the PAC of the demand subarray by 1, so it is effectively one PAC increment per demand ACT. Thus, CODA-E is 2x more efficient than CDP in terms of mitigations required to tolerate ColumnDisturb. 

Table~\ref{tab:codae} shows the rate of ACI for CDP and CODA-E.  For CDP, the rate remains 200\% as each activation increments two additional ACTR (one for each neighbor). For CODA-E, the rate depends on the number of bits in the PAC.  With a 4-bit PAC, the rate reaches the theoretical minimum for CODA-E (100\%). \new{We note that CODA-E is useful only for benign workloads and would not be effective for worst-case patterns that focus all activations on one subarray. }

\begin{table}[htb]
\centering
\begin{small}
\vspace{-0.05 in}
\caption{Rate of ACI for CDP and CODA-E}
\vspace{-0.15 in}
\label{tab:codae}
\begin{tabular}{cc}
\toprule
\textbf{Design} & \textbf{Rate of ACI} \\ \hline
\midrule
CDP                & 200\%  \\
CODA-E (1-bit PAC) & 154\%  \\
CODA-E (2-bit PAC) & 121\%  \\
CODA-E (3-bit PAC) & 104\%  \\
CODA-E (4-bit PAC) & 100\%  \\

\bottomrule
\end{tabular}
\vspace{-0.15 in}

\end{small}
\end{table}

\subsection{Impact on Performance Overheads}

As CODA-E reduces ACI, it has lower slowdown than CDP. Figure~\ref{fig:codaeperf} shows the slowdown of SALT, CDP, and CODA-E for TRHD=500. On average, SALT incurs a negligible slowdown of 0.3\%, CDP incurs a slowdown of 17.8\%, and CODA-E incurs a slowdown of 5.9\%.  Thus, CODA-E removes most of the slowdown of CDP.

Note that the slowdown of CODA-E can be 2x lower than that of CDP, as the slowdown for SALT depends on whether the activity is lower than what REF can handle. If SALT had half as many activations as REF can handle, the 3x activations (due to CDP) would incur significant overhead, whereas 2x activations (due to CODA-E) would incur no slowdown, so benefits are non-linear. 

\subsection{Impact on Security}

CODA-E has only a minor impact on the tolerated threshold.  With a 4-bit PAC, at most 15 ACI-based counter increments will remain pending and uncommitted to the ACTR. Table~\ref{tab:codaetrhd} shows the maximum activations to subarray under SALT (Subarray-Max, which is equal to 2*TRHD), the PAC-Max (15 for 4-bit counter), the total resulting activations with CODA-E (Total), and the percentage increase in the tolerated threshold.  Thus, the impact of CODA-E is 1.5\% at TRHD of 500 and 0.2\% at TRHD of 4K. 

\begin{table}[htb]
\centering
\begin{small}
\caption{Impact of CODA-E on Tolerated Threshold }
\vspace{-0.1 in}
\label{tab:codaetrhd}
\begin{tabular}{ccccc}
\toprule
\textbf{TRHD} & \textbf{Subarray-Max} & \textbf{PAC-Max} & \textbf{Total} & \textbf{Increase (\%)} \\ 
\midrule
500  & 1000 & 15 & 1015 & 1.5\% (508) \\
1K   & 2000 & 15 & 2015 & 0.8\%  (1.01K)\\
2K   & 4000 & 15 & 4015 & 0.4\% (2.01K)\\
4K   & 8000 & 15 & 8015 & 0.2\% (4.01K)\\
\bottomrule
\end{tabular}
\vspace{-0.05 in}
\end{small}
\end{table}

%% file: codaf.tex
\section{CODA-F: Fractional ACI-Based Increments}

Our second design for CODA, {\em CODA-F (Fraction)}, is based on the observation that ColumnDisturb is a much slower attack than Rowhammer.  For example, at TRHD=500, one could perform a Rowhammer attack within 50 microseconds (tRC of 50 ns, multiplied by 1000 activations, 500 each on two attack rows). Comparatively, ColumnDisturb takes several milliseconds to cause failures. For example, the ColumnDisturb paper~\cite{columndisturb} targets solutions for a duration of 8ms. This also means that the ColumnDisturb solutions have a much longer time (milliseconds) to perform the mitigation. 

CODA-F is based on the insight that a 1-to-1 increment between demand ACT and increment to neighboring subarrays is unnecessary.  We should perform ACI at a rate such that, even if the neighboring subarray receives no demand ACTs, it can still refresh all rows within the subarray {\em within the ColumnDisturb duration}. For example, an attacker may use a RowPress pattern on the aggressor row(s) to cause intra-subarray ColumnDisturb in the adjacent subarray.  We could consider each 500 ns duration of open time to be equivalent to one activation~\cite{saxena2024impress,salt}.  Then, over 8ms~\cite{columndisturb}, the attacker would need to perform 16K equivalent activations on the aggressor row(s).  If SALT is designed for a TRHD of 500, then SALT would refresh all rows in the demand subarray 16 times within 8ms.  The adjacent subarrays need to refresh all their rows only once over the 8ms period, so their counters need to be updated at a rate of only 1/16 per demand activation.  The respective {\em fractional rates (f)} for TRHD at 1K, 2K, and 4K are 1/8, 1/4, and 1/2, respectively.

\begin{figure*}[!htb]
    \centering
\includegraphics[width=7.2in]{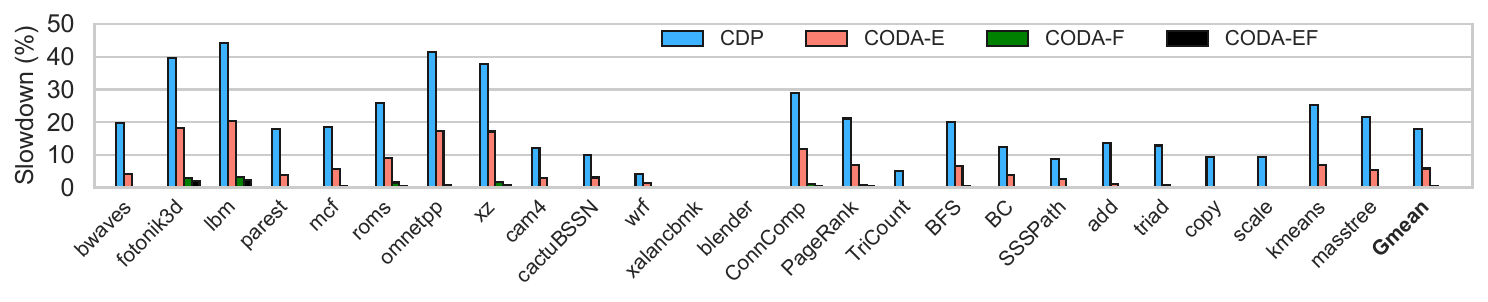}

\vspace{-0.15 in}
    \caption{Slowdown of CDP, CODA-E, CODA-F, CODA-EF at TRHD=500. The average slowdown of CDP is 17.8\%, of CODA-E is 5.9\%, of CODA-F is 0.6\%, and of CODA-EF is 0.3\% (Note: bars for CODA-F and CODA-EF are vanishingly small, hence not visible).}
\vspace{-0.1 in}
    \label{fig:codafperf}
\end{figure*}

\subsection{Overview and Design of CODA-F}

\begin{figure}[!htb]
    \centering
\includegraphics[width=1.6 in]{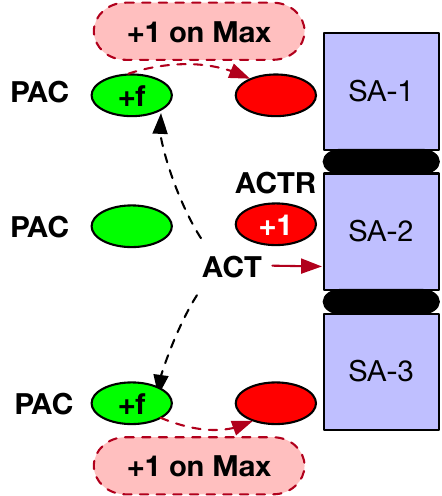 }
    \caption{Overview of CODA-F.  CODA-F performs fractional counter increments (by a value "f"=1/16 to 1/2) to PAC and increments ACTR if the PAC reaches the maximum value. }
   \vspace{-0.1 in}
    \label{fig:codaf}
\end{figure}

To implement CODA-F, we need to support a fractional increment of ACTR.  However, to avoid complexity (and to be synergistic with CODA-E), we first increment the ACI in a separate {\em Pending-Activation Counter (PAC)} by a sub-integer amount and write to the ACTR only if the PAC is 1 (or, alternatively, the maximum value).  Figure~\ref{fig:codaf} shows the overview of CODA-F.

Similar to CODA-E, CODA-F also provisions an auxiliary counter (PAC) with each subarray.  However, the representation of the value stored in PAC differs from that in CODA-E.  On activation of a demand subarray (e.g. SA-2), CODA-F increments the associated ACTR.  CODA-F also increments the PAC of the adjacent subarrays (e.g. SA-1 and SA-3) but by a smaller fractional amount (f), which is dictated by the ColumnDisturb duration and the tolerated TRHD (e.g. f=1/16 for TRHD=500 and ColumnDisturb duration of 8ms).  Thus, PAC stores the value in fixed-point format instead of integer.

If an ACI increment causes PAC to reach its maximum value, we reduce PAC by 1 and increment the ACTR of the associated subarray by 1.  Without loss of generality, we use a 4-bit PAC (which means it can count 15 fractional increments to reach the maximum value).  

\subsection{CODA-EF: Exploiting Synergy With CODA-E}

CODA-F by itself can reduce the overhead of ACI increments by a factor of 1/f.  However, we can combine CODA-E and CODA-F for even greater effectiveness.  The combined scheme, which we call {\em CODA-EF}, can be implemented by decrementing the PAC of the demand subarray by 1 (if the PAC is less than 1, it is reset to 0). 


\subsection{Impact on Rate of ACI}


Table~\ref{tab:codaf} shows the rate of ACI for CDP, CODA-E, CODA-F, and CODA-EF.  While CODA-E reduces ACI by 2x, CODA-F reduces ACI by 16x-2x.  When combined, the reduction is much larger, because if "f" is small (say 1/16), several activations are required for a PAC to reach 1, however, a single demand activation on the subarray resets PAC to zero. CODA-EF reduces ACI by 11.8x to 143x \new{(the average over all of our evaluated workloads)}. 


\begin{table}[htb]
\centering
\vspace{-0.1 in}
\begin{small}
\caption{Rate of ACI for CDP, CODA-E, CODA-F, and CODA-EF (number with "x" denotes relative reduction versus CDP)}
\vspace{-0.1 in}
\label{tab:codaf}
\begin{tabular}{lcccc}
\toprule
\textbf{TRHD} & \textbf{CDP} & \textbf{CODA-E} &  \textbf{CODA-F} & \textbf{CODA-EF}\\ \hline
\midrule
500 & 200\% & 100\% (2x) & 12.5\%  (16x)& 1.4\% (143x)\\
1K & 200\% & 100\%  (2x)& 25\% (8x) & 1.8\% (109x)\\
2K & 200\% & 100\%  (2x)& 50\%  (4x)& 3.6\% (56x)\\
4K & 200\% & 100\%  (2x)& 100\% (2x) & 17\% (11.8x)\\

\bottomrule
\end{tabular}
\vspace{-0.15 in}

\end{small}
\end{table}

\subsection{Impact on Performance Overheads}

Because CODA-F reduces the ACI rate more than CODA-E, it incurs a lower performance overhead than CODA-E. Figure~\ref{fig:codafperf} shows the slowdown of CDP, CODA-E, CODA-F, and CODA-EF at TRHD=500. The bar labeled Gmean shows the geometric mean slowdown.  On average, CDP incurs a slowdown of 17.8\%,  CODA-E of 5.9\%, CODA-F of 0.6\%, and the combination CODA-EF of 0.3\% (same as SALT without ColumnDisturb mitigation).  Thus, CODA-F removes nearly all of the slowdown caused by ColumnDisturb mitigation.


\subsection{Impact on Security}

CODA-F (and CODA-EF) has a negligible impact on the tolerated threshold.  With a 4-bit PAC, at most 15 ACI-based counter increments will remain pending. Table~\ref{tab:codaftrhd} shows the maximum activations to subarray under SALT (Subarray-Max, which is equal to 2*TRHD), the PAC-Max (depends on "f"), the total resulting activations with CODA-F (Total), and the percentage increase in the tolerated threshold.  Thus, the impact of CODA-F/CODA-EF is 0.1\% across all TRHD. 

\begin{table}[htb]
\centering
\begin{small}
\caption{Impact of CODA-F/CODA-EF on Tolerated Threshold }
\vspace{-0.1 in}
\label{tab:codaftrhd}
\begin{tabular}{ccccc}
\toprule
\textbf{TRHD} & \textbf{Subarray-Max} & \textbf{PAC-Max} & \textbf{Total} & \textbf{Increase (\%)} \\ 
\midrule
500  & 1000 & 1 & 1001 & 0.1\%  (501) \\
1K   & 2000 & 2 & 2002 & 0.1\%  (1.001K)\\
2K   & 4000 & 4 & 4004 & 0.1\% (2.002K)\\
4K   & 8000 & 8 & 8008 & 0.1\% (4.004K)\\
\bottomrule
\end{tabular}
\vspace{-0.15 in}
\end{small}
\end{table}

%% file: codag.tex
\clearpage

\section{CODA-G: Skip ACI for Intra-Gang Increments}

We observe that at thresholds of 1K or higher, we are likely to implement Ganged-SALT to reduce the stored overhead (e.g., at a 4K threshold, Ganged-SALT can be implemented with just 72 bytes of SRAM per bank, which is less than even some current TRR implementations). Ganged-SALT provisions a single {\em Activation Counter (ACTR)} over multiple subarrays of the gang, and all the rows in the gang are guaranteed to get refreshed before the gang receives 2*TRHD activations.  The multi-subarray nature of Ganged-SALT provides further opportunities to reduce ACI.  We call the CODA variant that exploits intra-gang inefficiency as {\em CODA-G}. 

\subsection{Overview and Design of CODA-G}

\begin{figure}[!htb]
    \centering
\includegraphics[width=1.8 in]{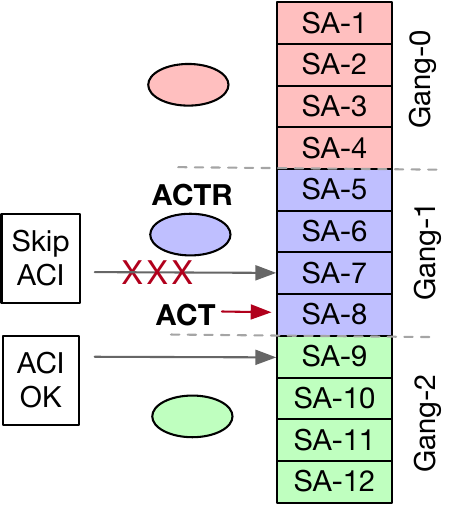 }
    \caption{Overview of CODA-G.  CODA-G exploits the multi-subarray granularity to skip ACI if the adjacent subarray is within the same gang. }
    \label{fig:codag}
\end{figure}

The key insight in CODA-G is to skip the ACI-based increment for the neighboring subarray when it is within the same gang.  We call this optimization {\em Gangskip}.  Figure~\ref{fig:codag} shows an overview of CODA-G.  Ganged-SALT uses four subarrays per gang (e.g., Gang-0 contains SA-1 to SA-4, and so on). Each gang has a single ACTR.  When a row is activated, the ACTR of the gang is incremented. For example, if SA-8 has an activation, the ACTR of Gang-1 is incremented.

A straightforward way to implement {\em ColumnDisturb Protection (CDP)} for Ganged-SALT is to perform ACI on the subarrays neighboring the accessed subarray. For example, on an activation to SA-8, we not only increment ACTR for SA-8 but also for SA-7 and SA-9. However, we note that the increment for SA-7 is unnecessary as it maps to the same ACTR as SA-8, for which we have already done an ACTR increment.  With CODA-G, we increment the adjacent subarray only if it is in a different gang (e.g., SA-9).  CODA-G skips the ACI activation-based counter increment if the adjacent subarray is in the same gang (e.g., SA-7). The insight is that the increment from demand activation is sufficient to protect all subarrays within the gang. With CODA-G, only a single ACI occurs if the activation targets a border subarray within the gang, and ACIs are skipped entirely for non-border subarrays (as both adjacent neighbors map to the same gang as the demand subarray).  Thus, CODA-G becomes even more efficient as the gang size increases. 

\subsection{Synergy of CODA-G with CODA-EF}

The advantage of CODA-G over CODA-E and CODA-F is that CODA-G can be implemented without incurring any additional storage, whereas CODA-E and CODA-F both require an additional counter (PAC).  However, both CODA-G and CODA-EF are complementary and, when combined, achieve greater reduction than either scheme alone. To implement CODA-G with CODA-EF, we implement CODA-EF at gang granularity and skip the ACI for neighboring subarrays within the same gang.  We refer to the combination of CODA-G and CODA-EF as {\em CODA-EFG}.

\subsection{Impact on Storage Overheads}

CODA-G incurs no additional storage overhead (the decision to skip is based solely on the address of the accessed subarray).  CODA-E, CODA-F, and CODA-EF require a 4-bit PAC counter with each subarray/gang.  Table~\ref{tab:storage} shows the storage overhead (SRAM bytes per bank) for Ganged-SALT, CODA-EF, CODA-G, and CODA-EFG. The storage overhead of Ganged-SALT with CODA remains quite small, ranging from 600 bytes at TRHD=500 to 88 bytes at TRHD=4K. 

\begin{table}[htb]
\centering
\begin{small}
\caption{Storage overhead (SRAM bytes per bank) for SALT, CODA-EF, Ganged-SALT, and CODA-EFG}
\vspace{-0.1 in}
\label{tab:storage}
\begin{tabular}{lcc|cc}
\toprule
\textbf{TRHD} & \textbf{\new{SALT}} & \textbf{\new{CODA-EF}} &  \textbf{\new{Ganged-SALT}} & \textbf{\new{CODA-EFG}}\\ \hline
\midrule
500 & \new{480B} & \new{608B}  & \new{N/A} & \new{N/A}   \\
1K & \new{544B} & \new{672B} & \new{256B} & \new{320B}  \\
2K & \new{576B} & \new{704B} & \new{136B} & \new{168B} \\
4K & \new{608B} & \new{736B} & \new{72B} & \new{88B} \\
\bottomrule
\end{tabular}
\vspace{-0.1 in}
\end{small}
\end{table}

\subsection{Impact on Rate of ACI}

With a gang of N subarrays and CODA-G standalone, the expected rate of ACI increments per demand activation is 2/N (so, 25\% for 8 subarrays per gang, which is relevant for TRHD=4K). The rate of ACI for CODA-EFG can be lower than the product of either scheme, standalone, as CODA-E uses demand activation to reset the PAC if it is below 1.  So, there is a non-linear benefit if the combination can frequently keep PAC below 1.

\begin{table}[htb]
\centering
\begin{small}
\caption{Rate of ACI for CDP, CODA-EF, CODA-G and CODA-EFG (number with "x" is relative reduction versus CDP)}
\vspace{-0.1 in}
\label{tab:codag}
\begin{tabular}{lcccc}
\toprule
\textbf{TRHD} & \textbf{CDP} & \textbf{CODA-EF} &  \textbf{CODA-G} & \textbf{CODA-EFG}\\ \hline
\midrule
500 & 200\% & 1.4\% (143x)  & 200\%  (1x)& 1.4\% (143x)  \\
1K & 200\% & 1.8\% (109x) &100\% (2x)& 0.23\% (869x) \\
2K & 200\% & 3.6\% (56x)& 50\%  (4x)&  0.15\% (1300x)\\
4K & 200\% & 17\% (11.8x)& 25\% (8x) & 0.16\% (1250x)\\

\bottomrule
\end{tabular}

\end{small}
\end{table}

Table~\ref{tab:codag} shows the rate of ACI per demand activation for CDP, CODA-EF, CODA-G, and CODA-EFG as TRHD is varied from 500 to 4K. CODA-G reduces the rate of ACI from 200\% (for CDP) to 25\% at TRHD of 4K. The combination of CODA-EFG is highly effective across all thresholds, as it uses CODA-G (which becomes more effective at higher TRHD) and CODA-EF (which becomes more effective at lower TRHD).  CODA-EFG provides a 143x-1300x reduction in the ACI rate compared to CDP.  Thus, CODA-EFG makes it possible to tolerate ColumnDisturb while incurring negligible (virtually zero) performance and power overheads.

\ignore{
\begin{figure*}[!htb]
    \centering
\includegraphics[width=7.2in]{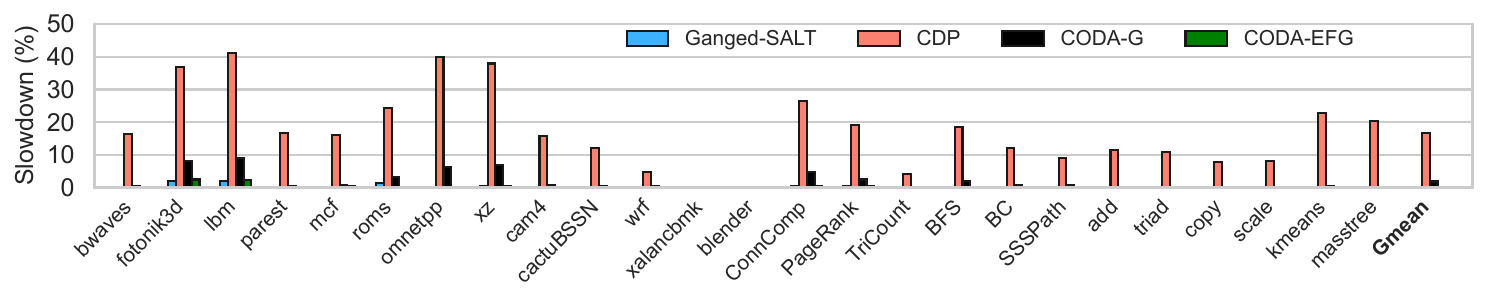}

    \caption{Slowdown of Ganged-SALT, CDP, CODA-G, and CODA-EFG for TRHD=4K. The average slowdown of Ganged-SALT is 0.3\%, CDP is 16.8\%, CODA-G is 1.9\%, and CODA-EFG is 0.3\% (Note: bars for Ganged-SALT and CODA-EFG are vanishingly small).}
\vspace{-0.1 in}
    \label{fig:codagperf}
\end{figure*}
}

\subsection{Impact on Performance Overheads}

As CODA-G and CODA-EFG significantly reduce the ACI, the performance overhead of mitigating ColumnDisturb becomes negligible.  Table~\ref{tab:codagperf} shows the slowdown for Ganged-SALT (without any ColumnDisturb mitigation) and Ganged-SALT implemented with CDP, CODA-G, and  CODA-EFG, as the TRHD is varied from 1K to 4K (gang sizes of 2 to 8 subarrays, respectively) and the target ColumnDisturb duration is 8ms.  On average, Ganged-SALT incurs a slowdown of only 0.3\%, whereas CDP increases it to 17.8\%. With CODA-G, the average slowdown ranges from 5.9\% (at TRHD=1K) to 0.9\%, with no additional storage overhead. In contrast, with CODA-EFG, the slowdown is identical to that of Ganged-SALT without any ColumnDisturb protection (average of 0.3\%). Thus, CODA-EFG can protect ColumnDisturb at  zero performance overhead.

\begin{table}[htb]
\centering
\begin{small}
\vspace{-0.1 in}
\caption{Average Slowdown of Ganged-SALT, CDP, CODA-G, and CODA-EFG at TRHD 1K-4K (ColumnDisturb of 8ms)}
\vspace{-0.1 in}
\label{tab:codagperf}
\begin{tabular}{ccccc}
\toprule
\textbf{TRHD} & \textbf{Ganged-SALT} &  \textbf{CDP} & \textbf{CODA-G} & \textbf{CODA-EFG}\\ \hline
\midrule

1K & 0.3\% & 17.7\% &5.9\% & 0.3\%  \\
2K & 0.3\% & 17.3\% & 1.9\%  &  0.3\% \\
4K & 0.3\% & 16.8\% & 0.9\%  &  0.3\% \\

\bottomrule
\end{tabular}
\vspace{-0.1 in}

\end{small}
\end{table}

\subsection{Impact on Security}

CODA-G exploits the multi-subarray nature of Ganged-SALT implementation to skip the update for the neighboring subarray that is within the same gang. Skipping this update has no impact on security, as the update is unnecessary: the ACTR of the gang is incremented due to demand activation, thereby protecting the entire gang.  When CODA-G is combined with CODA-EF, there is a minor impact of threshold from CODA-EF. Table~\ref{tab:codagtrhd} shows the effective TRHD with CODA-G and CODA-EFG for target TRHD of 1K to 4K. The impact of CODA-EFG is 0.1\% across all TRHD. 

\begin{table}[htb]
\centering
\begin{small}
\caption{Impact of CODA-G/CODA-EFG on Tolerated TRHD }
\vspace{-0.1 in}
\label{tab:codagtrhd}
\begin{tabular}{cccc}
\toprule
\textbf{Target-TRHD} & \textbf{CODA-G } & \textbf{CODA-EFG } & \textbf{Increase} \\ 
\midrule
1K   & 2000 & 2002 & 0.1\%  \\
2K   & 4000 &  4004 & 0.1\% \\
4K   & 8000 & 8008 & 0.1\% \\
\bottomrule
\end{tabular}
\end{small}
\end{table}

\subsection{Impact of ColumnDisturb Duration}

Similar to prior work~\cite{columndisturb}, we target a ColumnDisturb duration of 8ms. The ACI of CODA-E and CODA-G do not depend on the ColumnDisturb duration, so slowdowns remain unaffected. The ACI of CODA-F and CODA-EFG depends on the ColumnDisturb duration (as the "f" value is affected), so the slowdown can vary. Table~\ref{tab:codagperfcd} shows the slowdown for Ganged-SALT (without any ColumnDisturb mitigation) and Ganged-SALT implemented with CDP, CODA-G, and  CODA-EFG, as the TRHD is varied from 500 to 4K (at 500, Ganged-SALT degenerates to SALT) and the target ColumnDisturb duration is set to 4ms. The slowdowns of CDP and CODA-G remain the same. We note that CODA-G is not applicable at TRHD=500 because the gang contains only one subarray. Even at 4ms, CODA-EFG incurs 0\% additional slowdown compared to Ganged-SALT.

\ignore{
\begin{table}[htb]
\centering
\begin{small}
\vspace{-0.1 in}
\caption{Slowdown of Ganged-SALT, CDP, CODA-G, and CODA-EFG at TRHD 500-4K \new{(ColumnDisturb of 1/2/4/8 ms)}}
\vspace{-0.1 in}
\label{tab:codagperfcd}
\begin{tabular}{cccc|cccc}
\toprule
\textbf{TRHD} & \textbf{Ganged} &  \textbf{CDP} & \textbf{CODA-G} & \multicolumn{4}{c}{\new{\textbf{CODA-EFG}}} \\ \cmidrule(lr){5-8}
 & \textbf{SALT} &  & (1ms-8ms) & \new{\textbf{8ms}}  & \new{\textbf{4ms}} & \new{\textbf{2ms}} & \new{\textbf{1ms}}\\ 
\midrule

500 & 0.3\% & 17.8\% & N/A &  \new{0.3\%}   & \new{0.3\%}  & \new{0.4\%}  & \new{0.6\%} \\
1K  & 0.3\% & 17.7\% & 5.9\% & \new{0.3\%}   & \new{0.3\%}  & \new{0.4\%}  & \new{0.5\%} \\
2K  & 0.3\% & 17.3\% & 1.9\% & \new{0.3\%}   & \new{0.3\%}  & \new{0.4\%}  & \new{N/A} \\
4K  & 0.3\% & 16.8\% & 0.9\% & \new{0.3\%}   & \new{0.3\%}  & \new{N/A}  & \new{N/A} \\

\bottomrule
\end{tabular}
\end{small}
\end{table}
}

\begin{table}[htb]
\centering
\begin{small}
\vspace{-0.1 in}
\caption{Average Slowdown of Ganged-SALT, CDP, CODA-G, and CODA-EFG at TRHD of 500-4K (ColumnDisturb of 4ms)}
\vspace{-0.1 in}
\label{tab:codagperfcd}
\begin{tabular}{ccccc}
\toprule
\textbf{TRHD} & \textbf{Ganged-SALT} &  \textbf{CDP} & \textbf{CODA-G (4ms)} & \textbf{CODA-EFG (4ms)}\\ \hline
\midrule

500 & 0.3\% & 17.8\% & N/A & 0.3\%  \\
1K & 0.3\% & 17.7\% &5.9\% & 0.3\%  \\
2K & 0.3\% & 17.3\% & 1.9\%  &  0.3\% \\
4K & 0.3\% & 16.8\% & 0.9\%  &  0.3\% \\

\bottomrule
\end{tabular}

\end{small}
\end{table}

\new{We also consider scaling CODA to lower ColumnDisturb duration of 1ms-2ms.  However, in this regime, we must still ensure that the number of activations required for ColumnDisturb is more than what is needed for Rowhammer (e.g. if we treat row-open time of 500ns as one activation, then at 1ms duration, ColumnDisturb would need only 2K activations, so it is not meaningful to analyze TRHD of 2K or greater). So, for this analysis, we assume that row-open time of 250ns is equivalent to one activation (for our system, the adaptive page policy closes the page much earlier), and we conduct this analysis for TRHD of 1K and lower. }

\begin{table}[htb]
\centering
\begin{small}
\vspace{-0.05 in}
\caption{\new{Average Slowdown of Ganged-SALT, CDP, and CODA-EFG at TRHD of 500-1K (ColumnDisturb of 1-2ms)}}
\vspace{-0.1 in}
\label{tab:codagperfcdsmall}
\begin{tabular}{ccccc}
\toprule
\textbf{TRHD} & \textbf{Ganged-SALT} &  \textbf{CDP} & \textbf{\new{CODA-EFG}} & \textbf{\new{CODA-EFG} }\\ 
&  &  & \textbf{\new{2ms}} & \textbf{\new{1ms}}\\ \hline
\midrule

500 & 0.3\% & 17.8\% & \new{0.4\%}& \new{0.6\%}  \\
1K & 0.3\% & 17.7\% & \new{0.4\%} & \new{0.5\%}  \\

\bottomrule
\end{tabular}

\end{small}
\end{table}

%% file: rega.tex
\section{Efficiently Protecting REGA with CODA}
\label{sec:rega}


CODA can be implemented in any design that operates at subarray granularity, as such designs implicitly handle intra-subarray ColumnDisturb and can tolerate inter-subarray ColumnDisturb via ACI to neighboring subarrays.  Our solutions and insights for efficient ColumnDisturb mitigation are not limited to SALT.  In this section, we analyze another subarray-granularity mitigation, called {\em REGA (Refresh-Generating Activations)}~\cite{REGA_SP23}, and show how CODA can reduce the overheads for mitigating ColumnDisturb.

\subsection{Background on REGA}

\begin{figure}[!htb]
    \centering
    \vspace{-0.1 in}
\includegraphics[width=1.75 in]{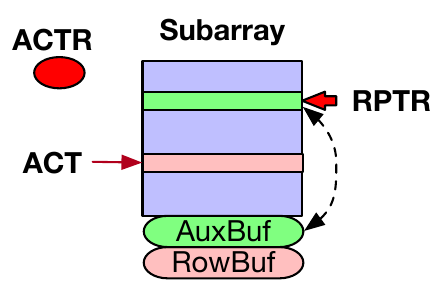 }
    \caption{Overview of REGA. REGA uses an auxiliary row buffer to concurrently refresh one/more rows on each ACT. }
   \vspace{-0.1 in}
    \label{fig:rega}
\end{figure}

The primary difference between SALT and REGA lies in the mechanism for obtaining mitigation resources.  While SALT uses ABO~\cite{JEDEC-PRAC} to obtain mitigation time on demand and incurs slowdowns due to ABO, REGA modifies the DRAM circuitry to obtain mitigation time for each activation.  Figure~\ref{fig:rega} shows the overview of REGA.

REGA modifies the subarray to have a second auxiliary row buffer {\em (AuxBuf)}.  On an activation, the accessed row is sensed using one row buffer, while the AuxBuf is used to concurrently refresh one (or more) rows (pointed by {\em RPTR}) from the subarray. The advantage of REGA is that, as mitigation occurs in the background, there is no time penalty if the refresh can be completed within the activation and precharge period of the accessed row. However, the disadvantage of REGA is that it modifies the DRAM circuitry, making it more difficult to adopt in commodity devices. Furthermore, it also exacerbates DRAM power consumption.

If one row is refreshed per ACT, and the subarray contains 512 rows, all rows are refreshed within 512 ACTs (tolerated TRHD of 256). To tolerate larger thresholds, REGA must refresh one row per several ACTs.  This can be achieved by having an {\em Activation Counter (ACTR)} per subarray.  ACTR is incremented on each activation. When ACTR reaches $N$, a refresh is performed concurrently with the ACT, and ACTR is reset.  For example, for TRHD=500, we need N=2, and for TRHD=4K, we need N=16.  This allows REGA to tolerate a higher threshold while incurring lower power overheads.


\subsection{REGA with ColumnDisturb Protection}

As REGA operates at subarray granularity, it tolerates intra-subarray ColumnDisturb. However, REGA cannot tolerate inter-subarray ColumnDisturb.  We can use CDP~\cite{salt} to protect REGA against ColumnDisturb, a design we call {\em REGA-CDP}. On activation, REGA-CDP not only increments the ACTR of the demand subarray but also that of the adjacent subarrays via ACI.  REGA-CDP incurs 3x the refresh-power overhead for mitigation compared to REGA.

\subsection{REGA-CODA: Overview and Design}

\begin{figure}[!htb]
    \centering
\includegraphics[width=1.75 in]{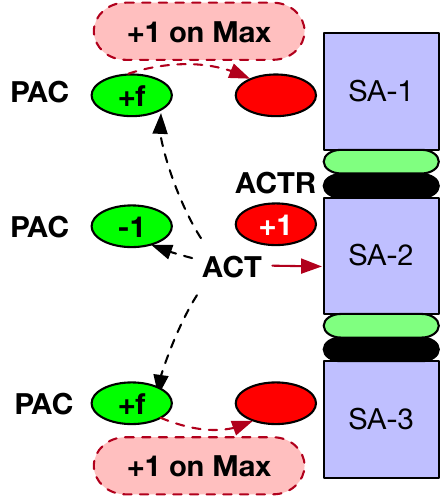 }
    \caption{Overview of REGA-CODA. REGA-CODA reduces/skips the ACI to the adjacent subarrays using insights from CODA-E (override by demand) and CODA-F (fraction). }
  \vspace{-0.1 in}
    \label{fig:regacoda}
\end{figure}

The power overhead of ColumnDisturb Protection for REGA can be reduced by applying CODA principles. Figure~\ref{fig:regacoda} shows the overview of such a {\em REGA-CODA} design. Each subarray has two counters, ACTR and PAC.  ACTR counts activations, and PAC counts delayed ACI from adjacent subarrays.  We use optimizations from both CODA-E (decrementing PAC on-demand activations to a subarray) and CODA-F (incrementing PAC by a fractional value "f" instead of one) to implement REGA-CODA.

On an activation, we increment the ACTR of the demand subarray.  If the ACTR reaches N (N=2, 4, 8, 16 for TRHD of 500, 1K, 2K, 4K, respectively), REGA refreshes a row and resets ACTR.  Similar to CODA-F, on activation, we also increment the PAC of the adjacent subarray by a value of "f" (f=1/16, 1/8, 1/4, 1/2 for TRHD of 500, 1K, 2K, 4K, respectively).  If PAC reaches its maximum value, we increment ACTR by 1 and reduce PAC by 1.  Similar to CODA-E, upon activation, we reduce the PAC of the demand subarray by 1 (and reset it to 0 if it was less than 1). The fractional increment and reset-on-demand-activations significantly reduce the rate of ACI.

\subsection{Impact on Rate of ACI}

REGA-CDP converts a single activation into three ACTR increments.  Thus, the rate of ACI is 200\%.  With REGA-CODA, this rate reduces significantly.  Table~\ref{tab:rega} shows the rate of ACI (per activation) for REGA-CDP and REGA-CODA.  The numbers in parentheses show the relative reduction compared to REGA-CDP. REGA-CODA reduces the ACI to 1.4\% (at TRHD=500) or 17\% (at TRHD=4K). Thus, REGA-CODA is much more efficient than REGA-CDP.

\begin{table}[htb]
\centering
\begin{small}
\caption{Rate of ACI for REGA-CDP and REGA-CODA (number with "x" shows relative reduction from REGA-CDP)}
\vspace{-0.1 in}
\label{tab:rega}
\begin{tabular}{ccc}
\toprule
\textbf{TRHD} & \textbf{REGA-CDP} & \textbf{REGA-CODA} \\ \hline
\midrule
500 & 200\% & 1.4\% (143x lower)  \\
1K & 200\% & 1.8\% (109x lower)  \\
2K & 200\% & 3.6\% (56x lower) \\
4K & 200\% & 17\% (11.8x lower)\\

\bottomrule
\end{tabular}

\end{small}
\end{table}

\subsection{Impact on Power Overheads}

For a mitigation rate of one (or fewer) row refreshes per ACT, REGA incurs no performance overhead, as the refresh happens concurrently with the activation. Thus, the performance of all REGA designs (REGA, REGA-CDP, REGA-CODA) is identical. The primary overhead of REGA is the additional power consumed for the mitigative refreshes.  This power overhead is proportional to the mitigation rate.

\begin{figure}[!htb]
    \centering
\includegraphics[width=2.75 in]{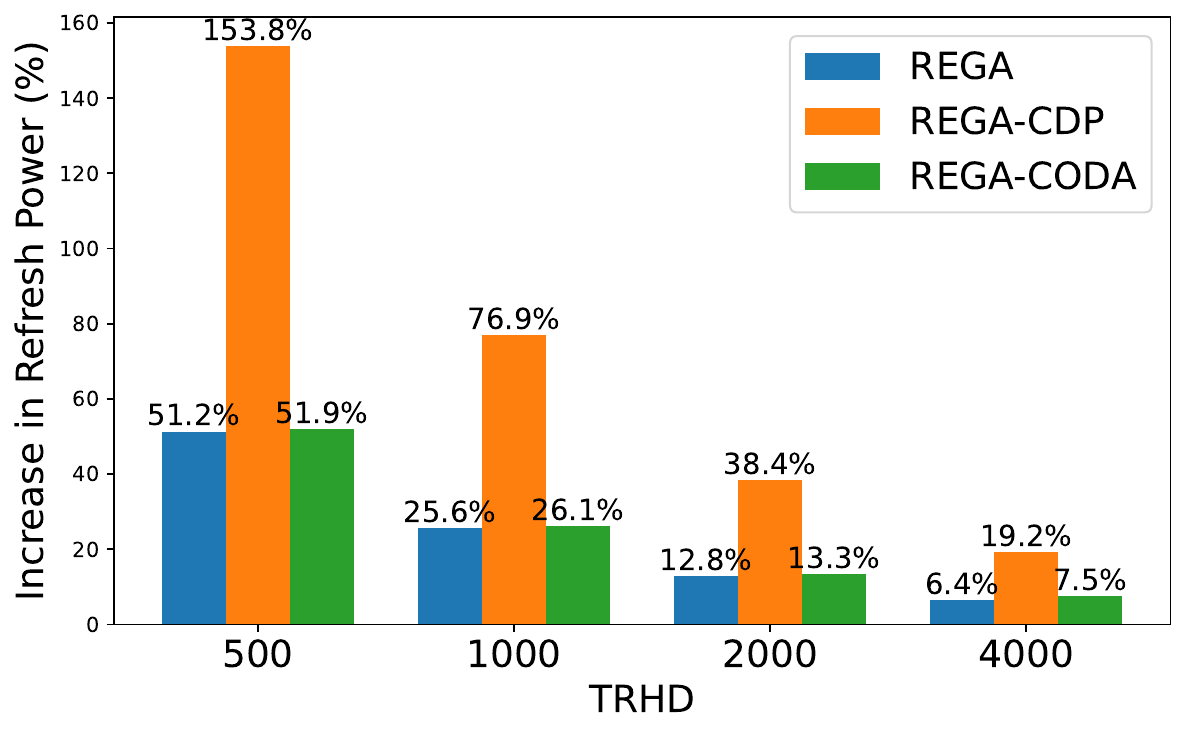 }
    \vspace{-0.1 in}

    \caption{Increase in refresh power with REGA, REGA-CDP, and REGA-CODA. REGA-CODA mitigates ColumnDisturb while incurring negligible power overheads. }
   \vspace{-0.1 in}
    \label{fig:regapower}
\end{figure}

Figure~\ref{fig:regapower} shows the increase in refresh power due to REGA, REGA-CDP, and REGA-CODA.  With REGA-CDP, the refresh power overhead triples, so the 51\% increase for REGA becomes 153\% with REGA-CDP. With REGA-CODA, the increase is negligible, increasing from 51\% to 52\%. Across all thresholds, the increase in power with REGA-CODA is within 1\% of that of REGA without any ColumnDisturb mitigation.  Thus, CODA can protect REGA against ColumnDisturb while incurring negligible power overheads.

%% file: related.tex
\section{Related Work}

As ColumnDisturb is a recently disclosed vulnerability (publicly released only five months ago), there are not many studies on mitigating ColumnDisturb. In this section, we describe work closely related to ColumnDisturb, as well as other Data-Disturbance Errors.

\subsection{ColumnDisturb Mitigation via Fast Refresh}

The ColumnDisturb paper~\cite{columndisturb} mentions two mitigations to limit ColumnDisturb to 8ms. First, reduce the DRAM refresh time from 32ms to 8ms, which incurs significant energy and performance overhead. Second, a scheme called {\em PRVR (Proactively Refreshing ColumnDisturb Victim Rows)}, which refreshes all rows in three subarrays once before the aggressor row is hammered or pressed enough times to induce a bit-flip. Unfortunately, the paper does not provide any design details for PRVR, such as the tracking mechanism, the number of activations targeted to 8ms, or the mitigation rate across the three subarrays to ensure that mitigations are completed within 8ms. The paper observes that PRVR still incurs approximately 30\% of the performance overhead and 25\% of the energy overhead of an 8ms refresh, indicating that the overhead of PRVR remains high. 


\begin{figure}[!htb]
    \centering
\includegraphics[width=3.25 in]{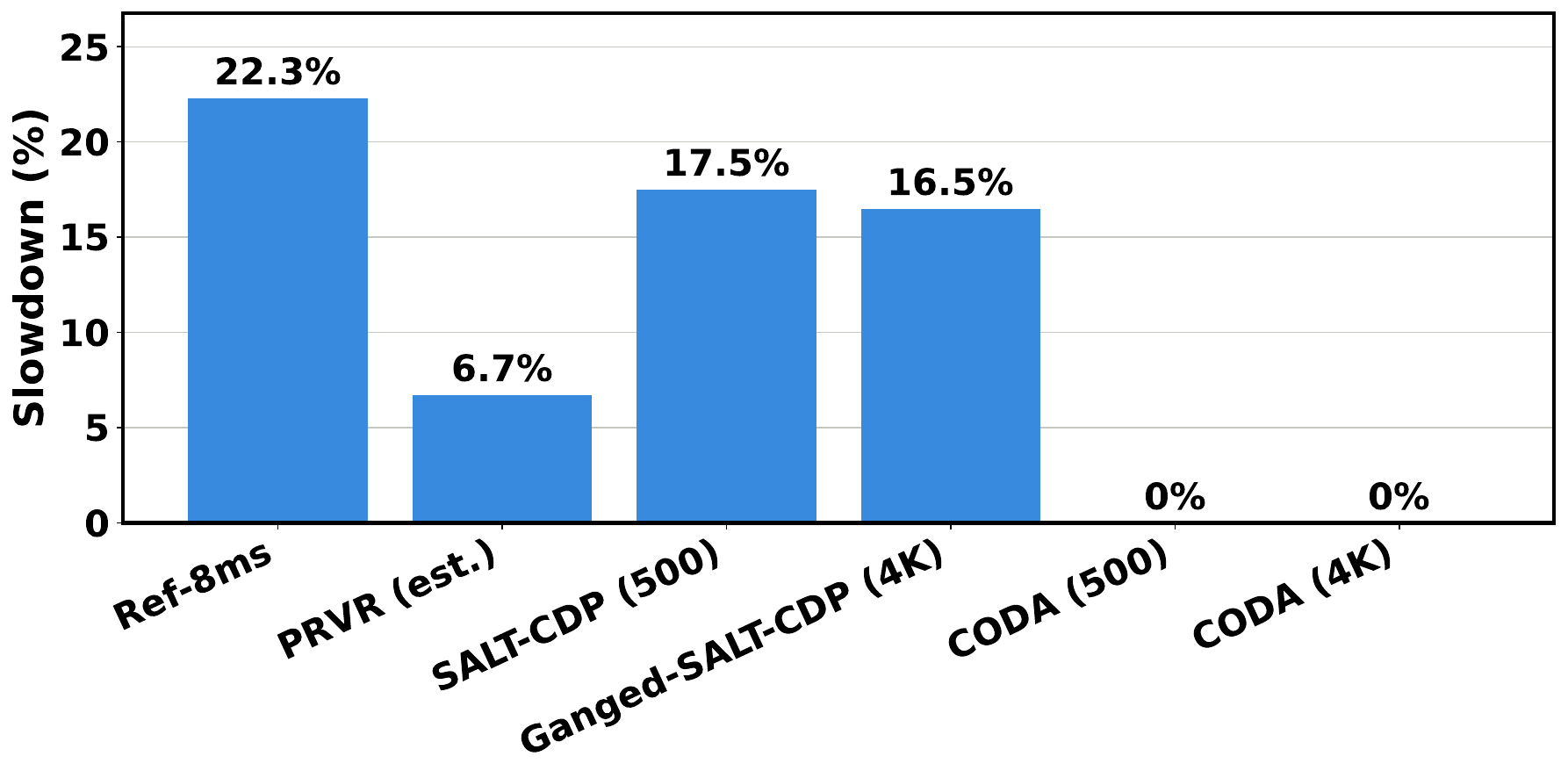 }
    \vspace{-0.15 in}
    \caption{Slowdown of Different ColumnDisturb Mitigations. CODA allows mitigations at 0\% slowdowns.  }
   \vspace{-0.15 in}
    \label{fig:prvr}
\end{figure}

Figure~\ref{fig:prvr} shows the slowdown of handling ColumnDisturb with Ref-8ms, PRVR (estimated), SALT-CDP-500 (TRHD=500), Ganged-SALT-CDP-4K (TRHD=4K), CODA-500 (TRHD=500), and CODA-4K (TRHD=4K). For SALT-based designs, we normalized the slowdown with respect to the SALT design without any ColumnDisturb protection. For PRVR, we estimate the slowdown as 30\% relative to Ref-8ms (as the paper does not provide sufficient design details). CODA-500 is built on top of SALT and uses CODA-EF.  CODA-4K is built on top of Ganged-SALT and uses CODA-EFG. An 8ms REF incurs a 22.3\% slowdown, and with PRVR (estimated), the slowdown remains 6.7\%.  CDP can incur 17\% slowdown, whereas CODA has 0\% slowdown for ColumnDisturb protection even at TRHD=500 (as it eliminates almost all of the ACI for neighboring subarrays). 

\subsection{Importance of Refresh Coordination}

We assume that both SALT and Ganged-SALT use Refresh Coordination~\cite{salt}  to avoid ABO.  CDP and CODA are also applicable to designs, such as {\em Silver-Bullet}~\cite{silverbullettr,silverbulletpatent}, that do not employ Refresh-Coordination. Such designs incur high overheads. Table~\ref{tab:sb} shows the slowdown of SALT(NR), SALT(NR) with CDP, and SALT(NR) with CODA-EF, where $NR$ denotes No-Refresh-Coordination.

\begin{table}[htb]
\centering
\begin{small}
\vspace{-0.1 in}
\caption{Slowdown of Designs with No-Refresh-Coordination (NR) for SALT, CDP, and CODA-EF.}
\vspace{-0.1 in}
\label{tab:sb}
\begin{tabular}{cccc}
\toprule
\textbf{TRHD} & \textbf{SALT(NR)} & \textbf{SALT(NR)+CDP} & \textbf{SALT(NR)+CODA-EF} \\ \hline
\midrule
500 & 13.2\% & 54.0\% & 13.5\%  \\
1K & 6.4 \% & 25.6\% & 6.6\%  \\
2K & 3.0\% & 12.0\%  & 3.2\% \\
4K & 1.5\% & 5.8\% & 1.9\%\\

\bottomrule
\end{tabular}
\vspace{-0.1 in}

\end{small}
\end{table}

SALT(NR) incurs 13\% at TRHD=500, and CDP increases it to 54\%, so the overheads of CDP are quite high.  With CODA-EF, the performance is within 0.5\% of the SALT(NR) design, which does not perform any ColumnDisturb mitigation.  Thus, CODA makes it practical to tolerate ColumnDisturb even for designs that do not employ Refresh-Coordination.

\subsection{Inefficacy of Row-Granularity Mitigations }

Typical hardware-based defenses for Rowhammer track aggressor rows and refresh a small number of victim rows adjacent to the aggressor row. Unfortunately, such row-granularity tracking cannot handle ColumnDisturb, as ColumnDisturb causes failures in rows hundreds of rows away from the aggressor row. For example, the state-of-the-art row-granularity mitigation proposal is PRAC+ABO from JEDEC~\cite{JEDEC-PRAC}. Several recent works, such as Chronus~\cite{canpolat2025chronus}, MOAT~\cite{qureshi2024moat}, and QPRAC~\cite{woo2025qprac}, use PRAC to securely tolerate Rowhammer. As their mitigations are limited to a small blast radius, they cannot handle failures in distant rows.

\subsection{Inefficacy of Rate-Limit Based Mitigations}

Prior works have tried to mitigate Rowhammer by limiting the number of times an aggressor row gets accessed to TRHD by employing some form of {\em Dynamic-Row-Migration} (such as {RRS}~\cite{saileshwar2022RRS}, {AQUA}~\cite{AQUA}, {SRS}~\cite{SRS}, Rubix~\cite{Rubix}, and SHADOW~\cite{ShadowHPCA23}) or explicitly enforcing rate limits (such as via Blockhammer~\cite{yauglikcci2021blockhammer}). Unfortunately, such designs are not effective at tolerating ColumnDisturb, as the attacker can spread the activations over hundreds of rows within a subarray, so each row would receive only dozens of activations and be within the permissible rate limits of these designs.

\ignore{
\subsection{Using Coding to Tolerate ColumnDisturb}

Another alternative to tolerate bit flips due to Data-Disturbance Errors, such as ColumnDisturb, is to use Error-Correction Codes (ECC). For example, {SafeGuard}~\cite{ali2022safeguard}, {CSI-RH}~\cite{csi},  {PT-Guard}~\cite{DSN23_PTGuard}, and Cube~\cite{twobirds}  modify the ECC codes to detect and correct Rowhammer failures, and these mechanisms would be equally effective for ColumnDisturb. Unfortunately, DDEs can cause more bit flips than the ECC limit allows; therefore, even in the presence of such solutions, uncorrectable failures and data loss can still occur.

}


%% file: conclusion.tex
\section{Conclusion}

As DRAM scales down, inter-cell interference increases, leading to new modalities of {\em Data-Disturbance Errors (DDEs)}.  {\em ColumnDisturb}, the newest form of DDEs, is particularly challenging to mitigate because it induces failures in rows hundreds of rows away from the aggressor row, and even in adjacent subarrays. ColumnDisturb can be mitigated by subarray-granularity mitigations, such as SALT, and using {\em Adjacent-Counter Increment (ACI)} for neighboring subarrays. Unfortunately, such a solution requires 3x mitigations and incurs significant overhead.  This paper proposes {\em CODA} to significantly reduce or eliminate the requirement of adjacent counter increments.  We propose three variants of CODA, and together they reduce the activation overheads of ColumnDisturb mitigation by 12x-1300x.  CODA makes it feasible to securely tolerate ColumnDisturb while incurring nearly zero performance overheads. 

\section*{Acknowledgments}

We thank the anonymous reviewers of MICRO-2026 for their feedback.  This work was funded, in part, by NSF grant 33304. 